# Surface atomic diffusion processes observed at milliseconds time resolution using environmental TEM

Wenpei Gao[1, 2], Jianbo Wu[1,2,3], Xiaofeng Zhang[4], Aram Yoon[1,2], J. Mabon[2], W. Swiech[2], W.L. Wilson[2], H. Yang[3] and Jian-Min Zuo[1, 2*]

[1]Dept of Materials Sci. and Eng., University of Illinois, Urbana-Champaign, IL 61801
[2]Seitz Materials Research Laboratory, University of Illinois, Urbana-Champaign, IL 61801
[3]Dept of Chemical and Biomolecular Eng., University of Illinois, Urbana-Champaign, IL 61801
[4] Hitachi High Technologies America Nanotechnology Systems Division, Pleasanton, CA 94588
*Corresponding author, jianzuo@illinois.edu

Significant progress has been made in spatial resolution using environmental transmission electron microscopes (ETEM), which now enables atomic resolution visualization of structural transformation under variable temperature and gas environments close to materials' real operational conditions (for a review, see ref [1]). Structural transformations are observed by recording images or diffraction patterns at various time intervals using a video camera [2] or by taking snap shots using electron pulses [3]. While time resolution at 15 ns has been reported using pulsed electron beams [3], the time interval that can be recorded by this technique is currently very limited. For longer recording, however, time resolution inside ETEM has been limited by electron cameras to ~1/30 seconds for a long time. Using the recently developed direct electron detection technology [4], we have significantly improved the time resolution of ETEM to 2.5 ms (milliseconds) for full frame or 0.625 ms for ¼ frames.

Here, we report the observation of Au atomic diffusion on a single crystal nanowire (NW) surface using the full frame readout at 2.5 ms. Experiment was carried out using Hitachi H-9500 ETEM with a $LaB_6$ emitter, operated at 300 kV. The instrument provides direct gas injection into sample chamber as well as gas injection through a specially designed sample holder. In imaging mode, the instrument is capable of resolving features in thin crystalline samples at 0.18 nm point to point. Sample is heated using a W filament by passing current [5]. For video recording, we use the direct electron detection camera (model K2-IS) made by Gatan (Pleasanton, CA). The camera has 3838x3710 pixels of 5 μm in size.

Figure 1 shows four image frames at 2.5 ms apart from a 300 seconds video recorded in bright field. The images focus on the tip of a silicon NW, where the Au catalyst after heating at 400 C° tilted from the tip to the side of the NW. When this occurs, Au can be seen diffuse away along the NW surface. At 0 ms (10min after the start of heating), Au diffusion was observed on the side of the NW as indicated by the arrow. At 2.5 ms, we saw further surface diffusion followed by a significant change in the wetting angle of Au nanoparticle at 5 ms, as well as image contrast change indicating significant amount of atomic movements. Image contrast associated with Au diffusion was observed over a distance of tens of nm along the NW within 7.5 ms. Further analysis shows a diffusion coefficient of approximately $10^{-8}\,cm^2/s$.

In summary, we have achieved a significant improvement in time resolution for in-situ environmental TEM. Preliminary results presented here demonstrate the capture of atomic diffusion process of Au on Si NW surface at 2.5 ms intervals. The time resolution is sufficient in revealing two processes, involving an initial wetting of Si NW surface by Au atoms followed by a change in Au nanoparticle morphology [6].

[6] We thank Cory Czarnik of Gatan Inc. for the development and installation of K2-IS camera. We also thank Profs. Hui Zhang and Jian Sha of Zhejiang University for providing Si NWs. The work reported here is supported by the NSF MRI Grant NSF DMR 12-29454 and NSF DMR 0449790. Jianbo Wu is partially supported by DOE BES under contract DEFG02-01ER45923.

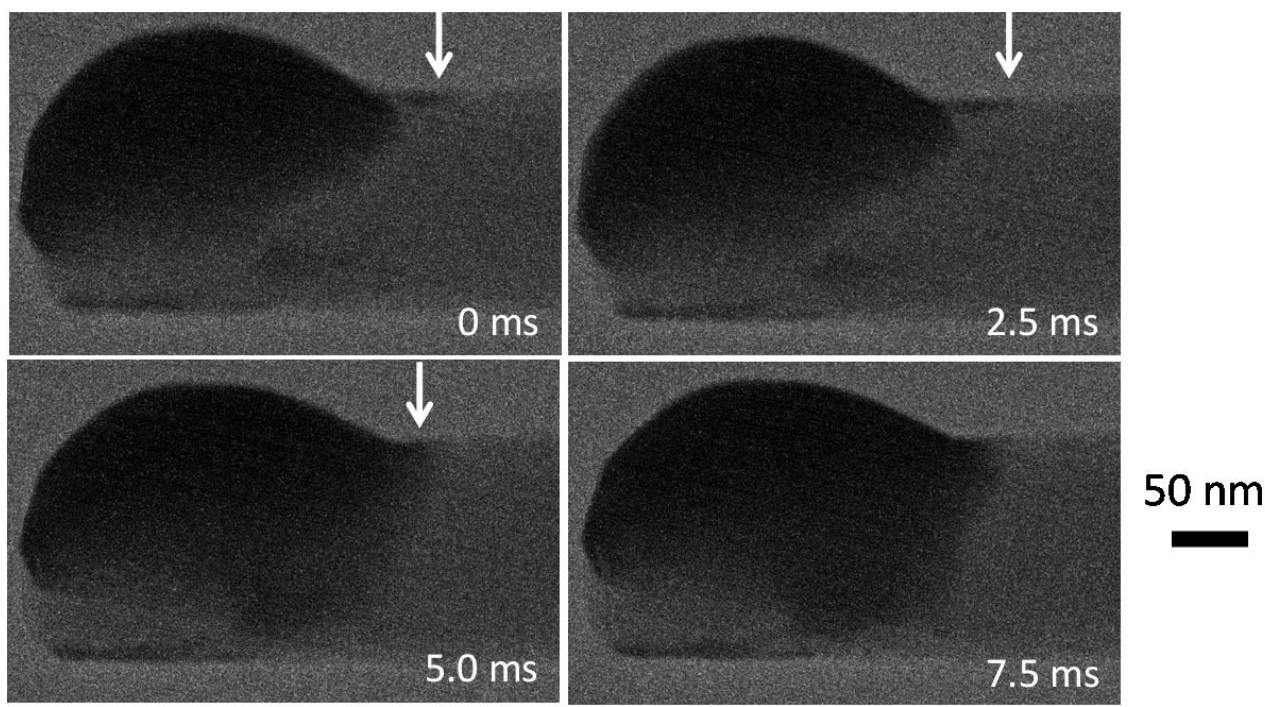

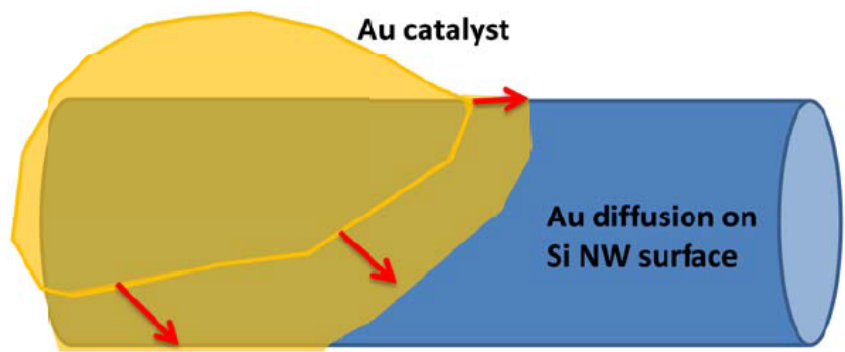

Figure 1 Top, bright field TEM images of the tip of SiNW taken at different time at 2.5 ms interval. The diffusion of Au atoms is indicated by the propagation of dark contrast starting from the tip along the NW surface vertically and horizontally. Below, a schematic illustration of Au diffusion on Si NW.